\documentstyle[12pt]{article}
\begin{document}

\baselineskip=15pt

\newcommand{\bc}{\begin{center}}
\newcommand{\ec}{\end{center}}
\newcommand{\be}{\begin{equation}}
\newcommand{\ee}{\end{equation}}
\newcommand{\bq}{\begin{eqnarray}}
\newcommand{\eq}{\end{eqnarray}}
\newcommand{\dd}{\frac{d^2k}{(2 \pi)^2}}
\newcommand{\dt}{\frac{d^3k}{(2 \pi)^3}}
\newcommand{\dtp}{\frac{d^3p}{(2 \pi)^3}}
\newcommand{\dq}{\frac{d^4k}{(2 \pi)^4}}
\newcommand{\dn}{\frac{d^nk}{(2 \pi)^n}}
\newcommand{\PLB}{{\it{Phys. Lett. {\bf{B}}}}}
\newcommand{\NPB}{{\it{Nucl. Phys. {\bf{B}}}}}
\newcommand{\PRD}{{\it{Phys. Rev. {\bf{D}}}}}
\newcommand{\AOP}{{\it{Ann. Phys. }}}
\newcommand{\MPL}{{\it{Mod. Phys. Lett. }}}

\begin{titlepage}

\vskip1in

\begin{center}
\LARGE{\bf Consistency Relations for an Implicit
$n$-dimensional Regularization Scheme}
\end{center}

\vskip1.0cm

\begin{center}
\Large{A. P. Ba\^{e}ta Scarpelli$^{\dagger}$  \,\, M.
Sampaio $^{\ddagger}$
\,\, M. C. Nemes$^{\dagger}$}\\ 
\end{center} 

\vskip1.0cm
\begin{center}
$\dagger$Federal University of Minas Gerais\\
Physics Department - ICEx\\
P.O. BOX 702, 30.161-970, Belo Horizonte MG - Brazil\\
\vskip1.0cm
$\ddagger$ University of Coimbra\\
Centre for Theoretical Physics\\
3004-516 Coimbra - Portugal\\
\end{center}
\vskip0.5cm
\begin{center}
{\it {carolina@fisica.ufmg.br,
msampaio@fisica.ufmg.br, scarp@fisica.ufmg.br}}
\end{center}
\begin{abstract}
\noindent
We extend an implicit regularization scheme to
be applicable in the $n$-dimensional space-time.
Within this scheme divergences involving parity
violating objects can be consistently treated without
recoursing to dimensional continuation. Special
attention is paid to 
differences
between integrals of the same degree of divergence,
typical of one loop 
calculations,
which are in principle undetermined. We show how to
use symmetries in 
order to fix
these quantities consistently. We illustrate
with examples in which  regularization plays a
delicate role in order to both corroborate and
elucidate 
the results in the literature for the case of CPT
violation in extended $QED_4$,
topological mass generation in $3$-dimensional gauge
theories, the Schwinger
Model and its chiral version.
\end{abstract}
\end{titlepage}


\section{Introduction and motivations}

The renormalizability  of the Standard Model (SM) of
particle physics underpins
its predictive power. Within perturbation theory, a
formal proof of renormalizability entreats a gauge
invariant regularization scheme. Whereas dimensional
regularization (DR) is known as the most powerful and
pragmatical method in the continuum space-time, care
must be exercised when dealing with theories whose
symmetry content depends on the space time dimension
such as chiral gauge theories and topological field
theories. In other words when parity violating objects
like $\gamma_5$ matrices or $\epsilon_{\mu_1 \mu_2
\ldots}$  tensors occur in the theory an appropriate
extension of DR must be performed since the properties
of these objects clash with the idea of analytical
continuation on the dimension of the space-time $n$.
This is the case of the electroweak sector of the SM 
\footnote{The only consistent framework in which the
renormalization of the electroweak SM in the continuum
$4$-dimensional space-time can be carried out to all
orders is algebraic renormalization within the BPHZ
formalism \cite{PIGUET}. However for practical
purposes it is rather involved as chiral and vector
gauge symmetries are broken in intermediate stages
what renders the calculations hard to handle.} as well
as Chern-Simons (CS) and CS-matter type of theories.
Yet such extension may be explicitly constructed,
namely the t' Hooft-Veltman dimensional continuation
(tHVDC), they are not unique and several
modifications were suggested \cite{EXT}. This, in
turn, may give rise to ambiguities (which we shall
discuss throughout this paper) and the appearance of
spurious anomalies. The latter is ultimately related
to the asymmetric definition of $\gamma_5$ when the
Dirac algebra is extended to $n$ dimensions
\cite{BURAS}. Although in one hand such shortcoming
may in principle be controlled by imposing the
Ward-Slavnov-Taylor identities order by order (and
introducing new finite counterterms), on the other
hand this turns the calculations significantly
cumbersome \cite{FERRARI}. Some very interesting views
on this subject have been recently presented
\cite{GAMMA5}.

Among the topological field theories in the
$3$-dimensional space-time,  the perturbative
computation of (pure) CS theories have applications to
both mathematics \cite{JONES} and physics \cite{CFT},
even though some exact results can be drawn
non-perturbatively \cite{NPCS}. 
When coupled to matter fields CS-matter theories are
no longer exactly solvable in general. Notwithstanding
they have a wide range of applications in condensed
matter physics (for a nice account see \cite{DUNNE},
\cite{WILCZECK} and references therein). In either
case the regularization ambiguities stemming from
their perturbative evaluation have been an everlasting
matter of debate \cite{CHAICHIAN-CHEN}, \cite{CSAMB}.
The third rank antisymmetric tensor in the CS
Lagrangian is just the three dimensional analogue of
the $\gamma_5$ in $4$-dimensional theory. A naive DR
cannot make the theory well defined \cite{ABBOTT}
whereas a so called dimensional reduction, that is the
evaluation of the entire antisymmetric tensor algebra
in $3$ dimensions while the loop momentum integrations
are performed in $n$ dimensions, can be shown to be
inconsistent \cite{CHAICHIAN-CHEN}. The most accepted
scheme is the BRS-invariant hybrid regularization
comprising a High Covariant Derivative (HCD) term
added in the Lagrangian (for instance a Yang-Mills
term $1/\Lambda^2 \,\, \mbox{tr} F^2$) \cite{SLAVNOV}
and the tHVDC. The former renders the model power
counting super-renormalizable and the remaining finite
number of diagrams which are left unregularized must
be regulated by the latter. The limits $n \rightarrow
3$ and $\Lambda \rightarrow \infty$ are well defined
\cite{MARTIN},\cite{RAO},\cite{CHENIR}
and should be taken in the end.  Some comments are in
order. If we consider CS theory
as a large topological mass limit of a topologically
massive Yang-Mills theory as it has been conjectured
by Jackiw \cite{JACKIWTMYM} then the Yang-Mills piece
of the Lagrangian is the natural candidate for the HCD
term. However, in principle one can construct higher
covariant derivative terms at will using covariant
derivatives
\cite{FALCETO}. In fact, as regards of the famous
shift/non-shift of the CS parameter \footnote{A
topological Ward identity 
constrains such shift to be an integer.}, a class of
(local, BRS-invariant) higher covariant derivative
regularizations were used (see Giavarini et al
\cite{CSAMB}): the shift depended on the large
momentum leading term of the regularized action  being
parity even or parity odd \footnote{This could well be
one the cases in which the radiative correction is
finite but undetermined since, non-perturbatively one
can only assert that the $\beta$-function vanishes to
all orders what does not discard a finite
correction.}. 

Another instance where the regularization ambiguities
play a delicate role is in non-renormalizable models,
often used as effective theories of QCD.
In such cases the regularization scheme is frequently
defined as part of the model 
and any parameters introduced by a specific choice
must be adjusted phenomenologically \cite{GHERKLE}.

Given the scenario above and the need to go to higher
orders in perturbation theory as the precision of the
experiments increase, it would be desirable to find a
regularization framework for diagrammatic computations
which preserved the advantages of DR without the need
to dimensionally continue
$\gamma_5$ or the antisymmetric tensor and/or
introduce HCD terms in the Lagrangian. 

Recently a step in this direction was taken. A
technique was proposed for the manipulation and
calculation of ($4$-dimensional) divergent amplitudes
in a way that a regularization need only to be assumed
implicitly \cite{OAC},\cite{AOMC}. The integrands are
algebraically manipulated until the infinities are
displayed in the form of basic divergent integrals in
the loop momenta which need not to be explicitly
evaluated in order to obtain the standard physical
results (they can be fully absorbed in the definition
of the renormalization constants). No dimensional
continuation is involved and a regulator needs only
implicitly to be assumed so to mathematically justify
the algebraic steps in the integrands of the divergent
integrals.

An important ingredient of this technique is a set of
consistency relations (CR) expressed by differences
between divergent integrals of the same degree of
divergence. 
In \cite{OAC} it was shown that such CR should vanish
in order to avoid ambiguities related to the various
possible choices for the momentum routing in certain
amplitudes involving loops, consistently with gauge
invariance. This is an important feature of DR and it
can be easily checked that the CR  are readily
fulfilled in the framework of DR. Alternatively and
more generically we can assign an arbitrary value to
such CR and let gauge invariance to determine its
value.  

Our purpose in this contribution is twofold. Firstly
to generalise this approach to be applicable for
theories defined in any dimension $n$ by deriving the
corresponding $n$-dimensional CR. This is very
important in order to both treat ambiguities related
to a particular choice of regularization and simplify
the loop calculations in dimensions other than four,
for instance in the CS-matter theories . Secondly, for
the purpose of illustration, we have selected examples
where different regularization schemes have somewhat
generated controversy in the literature.

This paper is organized as follows: In  section
\ref{S:RCN} we derive the CR for a $n$ dimensional
regularization and then we proceed to illustrate in
the context of theories defined in $4$-dimensions in
section \ref{S:QED4} for standard $QED_4$. In section
\ref{CPT} we revisit a well known example in
$4$-dimensions: the radiative generation of a CPT and
Lorentz violating Chern-Simons type term by
introducing  a term
$\bar{\psi}b\hspace{-2mm}/\gamma_5\psi$ in the
fermionic sector of QED. In section \ref{QED3} we
study the topological mass generation  in $3$-
dimensional QED and we analyse as an example in $2$
dimensions the Schwinger model and its chiral version
in section \ref{QED2}. Finally we draw our conclusions
and present some applications in which our scheme can
be useful.

\section{Consequences of momentum routing
independence}
\label{S:RCN}

Consider a one loop two-point function with two
vertices $\Gamma_i$ and $\Gamma_j$ and let $k$ be the
momentum running in the loop. In each propagator that
forms the loop we are allowed to add arbitrary
$4$-momenta, say $k_1$ and $k_2$, consistently with
the momentum-energy conservation. In \cite{OAC} it was
shown for $\Gamma_i=\Gamma_j=1$\, (SS),
$\Gamma_i=\Gamma_j=\gamma_5$ \, (PP),
$\Gamma_i=\gamma_\mu$ and $\Gamma_j=\gamma_\nu$ \,
(VV)
and $\Gamma_i= \gamma_\mu \gamma_5$ and
$\Gamma_j=\gamma_\nu \gamma_5$ \,(AA) that if these
amplitudes were to be independent of the arbitrary
momentum routing, that is to say, if they  were
translational invariant and consequently a shift the
momentum integration variable was allowed, then a set
of consistency relations (CR) between integrals of the
same degree of divergence had to hold in the sense
that the difference between the two integrals must
vanish. Such feature is manifest within DR (and hence
DR obeys the CR) but may not be fulfilled by other
gauge invariant regularizations. The existence of at
least two regularizations defined solely on the space
time dimension of the theory  that realise the CR was
also shown in \cite{OAC}. The same CR can be readily
derived by imposing translational invariance on the
(free) propagators of the theory. In this section we
derive the CR for an arbitrary space-time dimension
$n$.

For definiteness, consider the free fermionic Green´s
function
\be
S(x-x')=\int
\frac{d^nk}{(2\pi)^n}\frac{e^{ik(x-x')}}{k\hspace{-2.0mm}/-m}
\label{1}
\ee
and the corresponding ``translated" Green´s function
(i.e. a different representation of the same object)
\be
S_l(x-x')=\int 
\frac{d^nk}{(2\pi)^n}\frac{e^{i(k+l)(x-x')}}{k\hspace{-2.0mm}/
+ l\hspace{-2.0mm}/-m} \, ,
\label{2}
\ee 
where $l$ is an arbitrary momentum. Since we are
dealing with 
distributions, the
action of these objects require the definition of a
set of test 
functions on which
they act. It is straightforward to see that
translational invariance 
implies that
\be
\int S(x-x')j(x')d^nx'=\int S_l(x-x')j(x')d^nx'
\label{3}
\ee
Thus we conclude that $S_l(x-x')$ should be
independent of $l$:
\be
\frac{d}{dl}\int S_l(x-x')\eta(x')d^nx'=0 \, .
\label{4}
\ee
This condition guarantees that the generating
functional for the free 
theory 
$Z_0[\eta]=N\exp \bigg{\{}-i\int 
\overline{\eta}(x)S_l(x-y)\eta(y)d^nxd^ny \bigg{\}}$
does not depend on the particular Fourier
representation which has been 
used, provided
the test functions have the adequate physical
behavior. The same will hold true for the generating
functional of the interacting theory \cite{OAC}. Let
us take a closer look on the $l$ -dependence of the
Green's function
\be
\int S_l(x,y)\eta(y)d^ny=\int
d^ny\int\frac{d^np}{(2\pi)^n}\frac{e^{i(p+l)(x-y)}}
{p\hspace{-2.0mm}/+l\hspace{-2.0mm}/-m}\int
\frac{d^nq}
{(2\pi)^n}e^{iqy}\eta(q) \, .
\label{6}
\ee
One can now integrate over $y$ to obtain
\bq
\int
S_l(x,y)\eta(y)d^ny&=&\int\frac{d^np}{(2\pi)^n}\int\frac{d^nq}
{(2\pi)^n}\frac{e^{i(p+l)x}}{p\hspace{-2.0mm}/+l\hspace{-2.0mm}/-m}\eta(q)
\delta^n(q-p-l)
\nonumber 
\\
&=&\int\frac{d^nq}{(2\pi)^n}\frac{e^{i(p+l)x}}{p\hspace{-2.0mm}/+l\hspace{-2.0mm}/-
m}\eta(p+l) \, ,
\label{7}
\eq
which can also be conveniently rewritten in terms of
the translation 
operator as
\bq
\int
S_l(x,y)\eta(y)d^ny&=&\int\frac{d^np}{(2\pi)^n}e^{l^\mu\partial_\mu}\Bigg{(}
\frac{e^{ipx}}
{p\hspace{-2.0mm}/
-m}\eta(p)\Bigg{)} \nonumber \\
&=&\int\frac{d^np}{(2\pi)^n}\Bigg{\{}\frac{e^{ipx}}{p\hspace{-2.0mm}/-m}
\eta(p)\Bigg{\}}
+l^\mu\int\frac{d^np}{(2\pi)^n}\frac{\partial}{\partial
p^{\mu}}\Bigg{\{}\frac{e^{ipx}}{p\hspace{-2.0mm}/-m}\eta(
p)\Bigg{\}}+...
\label{8}
\eq
The first term on the RHS is the result which
expresses ``translational" 
invariance
as required by eq. (\ref{3}). All the other terms are
surface terms which,
provided $\eta(p)$ decays sufficiently fast as
required on physical 
grounds, should vanish identically. But on the
improper integrals, 
$S_l$ will act over
a distribuction,
\be
 \int S_l(x,y)D(x,y)d^nyd^nx,
\label{mud}
\ee
typically the delta function or products of particle
Greens function. 
So, we have
\be
\int S_l(x,y)D(x,y)d^nyd^nx=\int \frac{d^np}{(2\pi)^n}
 e^{l^\mu\partial_\mu}\Bigg{(}\frac{1}
{p\hspace{-2.0mm}/-m}D(p) \Bigg{)}.
\label{mud2}
\ee
For $D(x,y)=\delta(x-y)$, we have $D(p)=1$, and, for
instance, for the second 
term on the r.h.s. \be
l^{\mu}\int^{\Lambda}
\frac{d^np}{(2\pi)^n}\frac{\partial}{\partial
p^{\mu}}\Bigg{\{}\frac{1}{p\hspace{-2.0mm}/-m}\Bigg{\}}\,
.
\label{9}
\ee

At this point, since the integral is divergent, some
regulating
procedure must be adopted. Assume that the ultraviolet
divergent integrals in the momentum (say, $k$) are
regulated by the multiplication of the integrand by a
regularising function $G(k^2,\Lambda_i)$,
\be
\int_{k} f(k) \rightarrow \int \dn f(k)
G(k^2,\Lambda_i) \equiv \int^{\Lambda}_{k} f(k) \, ,
\ee
where $\Lambda_i$ are the parameters of a distribution
$G$ whose behavior for large $k$ renders the integral
finite. We shall only assume that such regulator is
even in $k$ and that the connection limit 
$\lim_{\Lambda_i \rightarrow \infty} G(k^2,\Lambda_i)
= 1$ is well defined. The latter will guarantee that
the value of the finite amplitudes will not be
affected by taking the limit.
Now the  integrand of (\ref{9})can be written as a
difference of two 
divergent integrals of the
same degree of divergence, namely
\be
\int^{\Lambda}
\frac{d^np}{(2\pi)^n}\frac{\partial}{\partial
p^{\mu}}\Bigg{\{}\frac{1}{\not{p}-m}\Bigg{\}}=\gamma_{\nu}\Bigg{\{}
\int^{\Lambda} 
\frac{d^np}{(2\pi)^n}\frac{g^{\mu\nu}}{p^2-m^2}-\int^{\Lambda}
\frac{d^np}{(2\pi)^n}\frac{2p^{\mu}p^{\nu}}{(p^2-m^2)^2}\Bigg{\}}
\label{10} 
\ee
If we vary the 
number of
Lorentz indices in the integrals, we obtain, for a
certain degree 
of divergence,
other relations in the higher orders of the expansion
(\ref{8}). Moreover the degree
of divergence of the 
integrals depend
on the dimension $n$. For example:
\subsection*{ 1+1 Dimensions: }
\be
\Delta_{\mu \nu}^0 \equiv \int^{\Lambda}
\frac{d^2k}{(2\pi)^2}\frac{g_{\mu\nu}}{k^2-m^2}-2\int^{\Lambda}
\frac{d^2k}{(2\pi)^2}\frac{k_{\mu}k_{\nu}}{(k^2-m^2)^2},
\label{11A}
\ee
\subsection*{ 2+1 Dimensions: }
\be
\Xi_{\mu \nu}^1 \equiv \int^{\Lambda}
\frac{d^3k}{(2\pi)^3}\frac{g_{\mu\nu}}{k^2-m^2}-
2\int^{\Lambda} 
\frac{d^3k}{(2\pi)^3}\frac{k_{\mu}k_{\nu}}{(k^2-m^2)^2},
\label{11}
\ee
\be
\Xi_{\mu \nu \alpha \beta}^1 \equiv
(g_{\mu\nu}g_{\alpha\beta}+g_{\mu\alpha}g_{\nu\beta}+g_{\mu\beta}g_{\nu\alpha})
\int^{\Lambda} \frac{d^3k}{(2\pi)^3}\frac{1}{k^2-m^2}
-8\int^{\Lambda} 
\frac{d^3k}{(2\pi)^3}\frac{k_{\mu}k_{\nu}k_{\alpha}k_{\beta}}{(k^2-m^2)^3},
\label{12}
\ee
etc..     
\subsection*{ 3+1 Dimensions: }
\be
\Upsilon_{\mu \nu}^2 \equiv \int^{\Lambda}
\frac{d^4k}{(2\pi)^4}\frac{g_{\mu\nu}}{k^2-m^2}-
2\int^{\Lambda} 
\frac{d^4k}{(2\pi)^4}\frac{k_{\mu}k_{\nu}}{(k^2-m^2)^2},
\label{13}
\ee
\be
\Upsilon_{\mu \nu}^0 \equiv \int^{\Lambda}
\frac{d^4k}{(2\pi)^4}\frac{g_{\mu\nu}}{(k^2-m^2)^2}-
4\int^{\Lambda} 
\frac{d^4k}{(2\pi)^4}\frac{k_{\mu}k_{\nu}}{(k^2-m^2)^3},
\label{14}
\ee
\be
\Upsilon_{\mu \nu \alpha \beta}^2 \equiv
(g_{\mu\nu}g_{\alpha\beta}+g_{\mu\alpha}g_{\nu\beta}+g_{\mu\beta}g_{\nu\alpha})
\int^{\Lambda} \frac{d^4k}{(2\pi)^4}\frac{1}{k^2-m^2}
-8\int^{\Lambda} 
\frac{d^4k}{(2\pi)^4}\frac{k_{\mu}k_{\nu}k_{\alpha}k_{\beta}}{(k^2-m^2)^3},
\label{15}
\ee
\be
\Upsilon_{\mu \nu \alpha \beta}^0 \equiv
(g_{\mu\nu}g_{\alpha\beta}+g_{\mu\alpha}g_{\nu\beta}+g_{\mu\beta}g_{\nu\alpha})
\int^{\Lambda}
\frac{d^4k}{(2\pi)^4}\frac{1}{(k^2-m^2)^2}
-24\int^{\Lambda} 
\frac{d^4k}{(2\pi)^4}\frac{k_{\mu}k_{\nu}k_{\alpha}k_{\beta}}{(k^2-m^2)^4},
\label{16}
\ee
etc. . Hence in order to assure momentum routing
independence, we have to set the
$\Delta$ 's, $\Xi$ 's, $\Upsilon$'s to vanish. A
simple illustration of this feature will be drawn in
section \ref{S:QED4}

It is interesting to notice that precisely the same
type of relations 
between divergent
integrals may appear in an $n$-dimensional theory in
connection with 
gauge invariance. In
order to show this let us consider a generic form for
the polarization tensor:
\be
\Pi_{\mu\nu}(k^2)=g_{\mu\nu}\Pi(0)+g_{\mu\nu}k^2\Pi_1(k^2)+k_{\mu}k_{\nu}\Pi_2(k^2).
\label{17}
\ee
Gauge invariance implies that
\be
k^{\mu}\Pi_{\mu\nu}(k^2)=0,
\label{18}
\ee
that is only true if $\Pi_{\mu\nu}(0)=0$. We can write
this, for the one 
loop calculation,
as
\be
\Pi_{\mu\nu}(0)=\int^{\Lambda} 
\frac{d^np}{(2\pi)^n}\frac{T_{\mu\nu}}{(p^2-m^2)^2},
\label{19}
\ee 
where
\be
T_{\mu\nu}=Ap^2g_{\mu\nu}+Bm^2g_{\mu\nu}+Cp_{\mu}p_{\nu},
\label{20}
\ee
and $A$, $B$ and $C$ are constants. Since
$\Pi_{\mu\nu}(0)=0$, we can 
suppose that the
integrand is a total derivative, and investigate if
there exist $A$, 
$B$ and $C$ which
satisfy the condition
\be
\frac{T_{\mu\nu}}{(p^2-m^2)^2}=\frac{\partial}{\partial
p^{\mu}}\Bigg{\{}\frac{Dp_{\nu}}{p^2-m^2}\Bigg{\}},
\label{21}
\ee
where $D$ is also a constant. After a simple algebra,
we conclude that $A=-B=D$ and 
$C=-2D$, so that
\be
\Pi_{\mu\nu}(0)=D\bigg{\{}\int^{\Lambda} 
\frac{d^np}{(2\pi)^n}\frac{g_{\mu\nu}}{p^2-m^2}-
2\int^{\Lambda} 
\frac{d^np}{(2\pi)^n}\frac{p_{\mu}p_{\nu}}{(p^2-m^2)^2}\bigg{\}}=0.
\label{GI}
\ee
In this case we may say that the same condition is
required to preserve both 
momentum routing independence and gauge invariance.
However in physical applications  we should privilege
the latter upon the former since there are examples in
which gauge invariance can only be attained at the
cost of adopting an especific momentum routing
\cite{JACTREI} namely when one axial vertex is
involved. We will come back to this issue in section
\ref{CPT}.

\section{$QED_4$}
\label{S:QED4}
In this section we illustrate our regularization
framework within $QED$ in $4$ dimensions so to compare
with well-known results as well as to gain some
insight
especially in the role played by an arbitrary routing
in the loop momentum of an amplitude in connexion with
the CR.

Consider the vacuum polarization tensor to one loop
order with arbitrary internal momentum routing
\be
\Pi_{\mu \nu }=\int \frac{d^4k}{\left( 2\pi \right)
^4}
\mbox{tr} \left\{ \gamma _\mu S(k+k_1)\gamma _\nu
S(k+k_2)\right\} ,
\label{tpqed}
\ee
where $S(k)$ is a usual half spin fermion propagator,
carrying momentum $k$. In order to make the arbitrary
momentum dependence more explicit, eq. (\ref{tpqed})
may be rewritten, after taking the trace over the
Dirac matrices, as
\bq
\Pi_{\mu \nu } &=&4 \Bigg(
\int^{\Lambda }\frac{d^{4}k}{(2\pi )^{4}}
\frac{2k_{\mu }k_{\nu
}}{[(k+k_{1})^{2}-m^{2}][(k+k_{2})^{2}-m^{2}]}  \\
&&+(k_{1}+k_{2})_{\nu }\int^{\Lambda
}\frac{d^{4}k}{(2\pi )^{4}}\frac{k_{\mu
}}{[(k+k_{1})^{2}-m^{2}][(k+k_{2})^{2}-m^{2}]} 
\nonumber \\
&&+(k_{2}+k_{1})_{\mu }\int^{\Lambda
}\frac{d^{4}k}{(2\pi )^{4}}\frac{k_{\nu
}}{[(k+k_{1})^{2}-m^{2}][(k+k_{2})^{2}-m^{2}]} 
\nonumber \\
&&+ \left( k_{2\mu }k_{1\nu }+k_{1\mu }k_{2\nu
}\right) \int^{\Lambda }
\frac{d^{4}k}{(2\pi
)^{4}}\frac{1}{[(k+k_{1})^{2}-m^{2}][(k+k_{2})^{2}-m^{2}]
}\Bigg)   \nonumber \\
&&- 2\, g_{\mu \nu }\Bigg( \int^{\Lambda
}\frac{d^{4}k}{(2\pi)^{4}}\frac{1}{[(k+k_{1})^{2}-m^{2}]}+
\int^{\Lambda
}\frac{d^{4}k}{(2\pi)^{4}}\frac{1}{[(k+k_{2})^{2}-m^{2}]}
\nonumber \\
&& -(k_{1}-k_{2})^{2}\int^{\Lambda
}\frac{d^{4}k}{(2\pi)^{4}}
\frac{1}{[(k+k_{1})^{2}-m^{2}][(k+k_{2})^{2}-m^{2}]}\Bigg)
. 
\label{tpqed2} 
\eq
Now we manipulate algebraically the integrands until
the external momentum dependence appears solely in
finite integrals by means of the identity 
\be
\frac 1{[(k+k_i)^2-m^2]}=\sum_{j=0}^N\frac{\left(
-1\right) ^j\left( k_i^2+2k_i\cdot k\right) ^j}{\left(
k^2-m^2\right) ^{j+1}}
+\frac{\left( -1\right) ^{N+1}\left( k_i^2+2k_i\cdot
k\right) ^{N+1}}{\left(
k^2-m^2\right) ^{N+1}[ \left( k+k_i\right) ^2-m^2] },
\label{rr}
\ee
$i=1,2$ and $N$ is such that the last term in
(\ref{rr}) is finite under integration over $k$
,\cite{AOMC}. After some straightforward
algebra, we can cast (\ref{tpqed2}) in the form:
\bq
\Pi_{\mu \nu } &=& \tilde\Pi_{\mu \nu } +
4\Bigg(\Upsilon^2_{\mu\nu}-\frac{1}{2}(k_1^2+k_2^2)\Upsilon^0_{\mu
\nu}
+\frac{1}{3}(k_{1}^{\alpha}k_{1}^{\beta}+k_{2}^{\alpha}k_{2}^{\beta}
+k_{1}^{\alpha}k_{2}^{\beta})
\Upsilon^0_{\mu \nu \alpha 
\beta} \nonumber \\ &-&
(k_1+k_2)^{\alpha}(k_1+k_2)_{\mu}\Upsilon^0_{\nu
\alpha} -\frac{1}{2}(k_1^{\alpha}k_1^{\beta}+k_2^{\alpha}k_2^{\beta})g_{\mu \nu}
\Upsilon^0_{\alpha \beta}  \Bigg)\,\,\,\, \mbox{where}
\label{qedvp}
\eq
\vspace{-0.5cm}
\bq
\tilde\Pi_{\mu \nu } &=& \frac{4}{3} \Big(
(k_1-k_2)^2g_{\mu
\nu }-(k_1-k_2)_\mu (k_1-k_2)_\nu \Big) \times 
\nonumber \\
&\times& \Bigg( I_{log}^\Lambda (m^2) - \frac i{(4\pi
)^2} \Bigg( \frac
13+\frac{(2m^2+\left( k_1-k_2\right) ^2)}{\left(
k_1-k_2\right) ^2}
Z_0(\left( k_1-k_2\right) ^2;m^2)\Bigg) \Bigg) \, , 
\eq
and the $\Upsilon$'s are the CR defined in (\ref{13})
- (\ref{16}), 
\be
I_{log}^\Lambda (m^2)= \int^\Lambda \frac{d^4k}{(2
\pi)^4}\frac{1}{(k^2-m^2)^2} \,\,\, \mbox{and} \,\,\, 
Z_0 (p^2;m^2) = \int_0^1 dz \, \ln \Big(
\frac{p^2z(1-z)-m^2}{-m^2}\Big)
\label{ilogz0}
\ee
It is clear from (\ref{qedvp}) that in order to
eliminate the ambiguous terms and to respect
the Ward identities $(k_1-k_2)^\mu \Pi_{\mu \nu }=
(k_1-k_2)^\nu  \Pi_{\mu \nu }=0$, we must set all the
$\Upsilon$'s to zero. Therefore we obtain  the usual
result for the vacuum polarization tensor.

Now let us adopt the particular routing $k_1=p$ and
$k_2=0$ and hence let the value of the CR to be
arbitrary, namely 
\bq
\Upsilon^0_{\mu\nu\alpha\beta}&=& c
(g_{\mu\nu}g_{\alpha\beta}+g_{\mu\alpha}
g_{\nu\beta}+g_{\mu\beta}g_{\alpha\nu})\, , \nonumber 
\\
\Upsilon^0_{\mu\nu} &=& a g_{\mu\nu}\, . 
\eq
Thus we have
\be
p^{\mu}\Pi_{\mu\nu}=4\left\{p^{\mu}\Upsilon^2_{\mu\nu}+
(c-2a)p^2p_{\nu} \right\}
\ee
from which we see that gauge invariance is implemented
for the choice 
\be
\Upsilon^2_{\mu\nu}=0 \,\, , \,\, c=2a
\ee
This will be important for the discussions in section
\ref{CPT}.  

Notice that at this point we can compare our result
with any sound regularization procedure, for instance
DR by an explicit computation of $I_{log}^\Lambda
(m^2)$. However, as far as the physical content is
concerned, one needs not to do so. For instance
consider the calculation of the $\beta$-function to
one loop order. We add the usual counterterm to define
$\tilde{\Pi}_{\mu \nu R}= \tilde{\Pi}_{\mu \nu} +
(q_\mu q_\nu - q^2
g_{\mu \nu})(Z_3 - 1)$, $A^\mu = Z_3^{1/2} A^\mu_R$
and $q = k_1 - k_2$. The Callan-Symanzik   
$\beta$-function  can be written as
$$
\beta = e_R \frac{\partial }{\partial \ln
\Lambda}\Big( \ln Z_3^{1/2}(e,\Lambda /m)\Big)
$$   
We may choose the renormalization constant such that
$(Z_3 - 1) = \frac{4}{3} i I_{log}^\Lambda (m^2)$
(which amounts to a subtraction at $q=0$), to get the
well known one loop result $\beta = 1/(12 \pi^2)$
($e_R=1$) where we used that $\partial I_{log}^\Lambda
(m^2)/\partial m^2 = -i/((4 \pi)^2 m^2)$. In
\cite{AOMC} we also calculate the $\beta$-function of
$\varphi^4_4$-theory to two loop order within this
approach.

\section{Induced Lorentz and CPT symmetry breaking in
extended $QED_4$}
\label{CPT}

While introducing a Chern-Simons term 
\be
{\cal{L}}_c = \frac{1}{2} c_\mu \epsilon^{\mu \nu
\lambda \rho} F_{\nu \lambda}A_\rho \, , \,\,\,
\mbox{$c_\mu$ being a constant $4$-vector, }
\label{lcptbt}
\ee
to violate Lorentz and CPT symmetries in conventional
$QED_4$ \cite{JACKCPT1}undergoes stringent theoretical
and experimental bounds
\cite{ANDRIA},\cite{KOSTE},\cite{COLEMAN}, there have
been investigations on possible extensions of the
Standard Model which could give rise to Lorentz and
CPT violation \cite{COLLA}. A natural question is
whether the term expressed in equation (\ref{lcptbt})
could be generated radiatively when Lorentz and CPT
violating terms occur in other parts of a larger
theory. For instance, many authors have exploited the
possibility of such term being induced by introducing
an
explicit Lorentz and CPT violating term $b_\mu
\bar{\psi}\gamma_\mu \gamma_5 \psi$ in the fermionic
sector of standard $QED_4$
\cite{JACKOS},\cite{CHUNGOH},\cite{CHENDR},\cite{PEREZ},\cite{CHUNGJ}.
In fact, a meticulous work by W. F. Chen and G.
Kunstatter \cite{CPTLS} seems to rule out such
particular extension  by studying its effect on the
calculation of the lambda-shift and on the anomalous
magnetic moment. Hence it would not constitute  a
physically plausible source of radiatively induced
terms like (\ref{lcptbt}). However, since the issue
here is the regularization dependence which is
involved in the radiative  correction, such
calculation serves as a perfect laboratory for
examining our framework.  

Consider the modified fermionic sector of $QED_4$ 
\be
{\cal{L}}_{fermion} = \bar{\psi}(i \partial
\hspace{-2.0mm}/ - A\hspace{-2.0mm}/ -
b\hspace{-2.0mm}/ \gamma^5 -m )\psi \,
\ee
where $b_\mu$ is a constant $4$-vector which selects a
specific direction in space-time and therefore the
gauge invariant term $\bar{\psi}b \hspace{-2.2mm}/
\gamma^5 \psi$ explicitly violates CPT and  Lorentz
symmetries. The quantity of interest for deciding
whether (\ref{lcptbt}) is radiatively generated is the
$O(A^2)$ part of the extented effective action 
\be
\Gamma_{ext} (A) = -i {\mbox{ln  det}}(i \partial
\hspace{-2.0mm}/ - A\hspace{-2.0mm}/ -
b\hspace{-2.0mm}/ \gamma^5 -m )
\ee
from which the coefficient $c_\mu$ is determined from
$b_\mu$. To lowest order in $b$, this corresponds
diagramatically to a triangle graph composed of two
vector currents and one axial vector current with
zero-momentum transfer between the two vector gauge
field vertices. Hence we can write generically that 
\be
\Gamma^2(A) \equiv 
\int\frac{d^4p}{(2\pi)^4}A^{\mu}(-p)\Pi_{\mu\nu}(p)A^{\nu}(p)
\, , 
\ee
with $\Pi_{\mu\nu}(p) \sim b_\alpha \Gamma^{\mu \nu
\alpha}(p, -p)$. Now as it was discussed in
\cite{JACKIWFU}, $\Gamma^{\mu \nu \alpha}(p, -p)$ is
undetermined by an arbitrary parameter $a$, namely
\be
\Gamma^{\mu \nu \alpha}(p, -p) \sim \Gamma^{\mu \nu
\alpha}(p, -p) + 2 i a \epsilon^{\mu \nu \alpha \beta}
p_\beta \, ,
\label{indet}
\ee
which cannot be fixed by requiring transversality of
$\Gamma^{\mu \nu \alpha}$. This is in constrast with
the famous triangle anomaly and it is essentially
because in our case the axial vector carries zero
momentum. Moreover there is no anomaly in the axial
current conservation law in this case. The
indetermination expressed in (\ref{indet}) (and
therefore in $c_\mu$) is apparent in the $b_\mu$
perturbative approach \cite{CHENDR},\cite{CHUNGPI}. 
However, following Jackiw \cite{JACKOS} one can also
carry out a non-perturbative calculation by employing
the $b_\mu$-exact propagator 
\be
S'(k)=
\frac{i}{ik\hspace{-2.3mm}/-m-b\hspace{-2.0mm}/\gamma_{5}}
\label{exprop}
\ee
which appears to lead to a definite unambiguous result
\cite{JACKOS},\cite{PEREZ},\cite{CHUNGJ}.
Before proceeding to study this problem within our
approach, a few comments are in order following
\cite{JACKOS}. Because the axial current $j_\mu^5(x)
\equiv \bar{\psi}(x)\gamma_\mu \gamma^5 \psi(x)$ does
not couple to any physical field but $b_\mu$, physical
gauge invariance is achieved provided that $j_\mu^5$
is gauge invariant at zero $4$-momentum. This is
equivalent to state that it is the
\underline{integrated} quantity $\int d^4x j_\mu^5(x)$
which is gauge invariant, in consonance with the fact
that the induced quantity which we seek
(\ref{lcptbt})is not gauge invariant while its
spacetime integral is. Hence, according to Jackiw
\cite{JACKIWFU} any regularization which enforces
gauge invariance at all momenta will render a
vanishing result for $c_\mu$ such as Pauli-Villars
regularization \cite{COLLA}; As for DR, there is not a
unique prescription to work within this scheme in the
presence of a $\gamma_5$ matrix and one has as many
results as alternative continuation prescriptions. 

We believe that within our scheme, which preserves the
characteristics of the theory as much as possible, one
has a good setting to study this problem. For this
purpose we illustrate it for both the non-perturbative
and the perturbative in $b_\mu$ treatments. 
We start by calculating the induced term in the
non-perturbative in $b_\mu$ scheme \cite{JACKOS}. The
exact propagator (\ref{exprop})can be separated as
\be
S'(k)=S_F(k)+S_b(k),
\ee
where $S_F(k)$ is the usual free fermion propagator
and
\be
S_b(k)=
\frac{1}{ik\hspace{-2.0mm}/-m-b\hspace{-2.0mm}/\gamma_{5}}
b\hspace{-2.0mm}/ 
\gamma_{5}S_F(k).
\ee
whereas the vacuum polarization tensor can be
generically written as in \cite{JACKOS}
\be
\Pi^{\mu \nu}=\Pi^{\mu \nu}_0 + \Pi^{\mu \nu}_b +
\Pi^{\mu \nu}_{bb}.
\label{tva}
\ee
We are concerned about the second term which is
linearly divergent and 
thus  it can be
responsible for a momentum-routing ambiguity.
Explicitly we have
\be
\Pi^{\mu \nu}_b(p)=\int\frac{d^4k}{(2\pi)^4} \mbox{tr}
\left\{\gamma^{\mu}S_F(k)
\gamma^{\nu}S_b(k+p) + \gamma^{\mu}S_b(k)
\gamma^{\nu}S_F(k+p) \right\}.
\ee
To lowest order in $b_\mu$, we can replace $S_b(k)$
with
$-iS_F(k)b\hspace{-2.0mm}/\gamma_5 S_F(k)$, so that 
$\Pi^{\mu \nu}_b=\Pi^{\mu \nu \alpha}b_{\alpha}$, 
with
\bq
\Pi^{\mu \nu \alpha}(p)&=& -i
\int\frac{d^4k}{(2\pi)^4} \mbox{tr} 
\left\{\gamma^{\mu}S(k)
\gamma^{\nu}S(k+p)\gamma^{\alpha}\gamma_5 S(k+p)
 + \gamma^{\mu}S(k)\gamma^{\alpha}\gamma_5 S(k )
\gamma^{\nu}S(k+p) 
\right\} \nonumber
 \\
&\equiv&-\left\{I_1^{\mu \nu \alpha} + I_2^{\mu \nu
\alpha} \right\}
\eq
We shall calculate the two  integrals separetely,
without doing a shift for the sake of clarity. The
ambiguities in momentum routing discussed by 
Jackiw will be made
explicit in the relations between divergent integrals
that will 
appear. After taking the trace over the Dirac matrices
we have
\be
I_1^{\mu \nu \alpha}= \int\frac{d^4k}{(2\pi)^4}
\frac{{\cal{N}}_1^{\mu \nu 
\alpha}}
{[k^2-m^2][(k+p)^2-m^2]^2} \,\, {\mbox{and}}
\ee
\be
I_2^{\mu \nu \alpha}= \int\frac{d^4k}{(2\pi)^4}
\frac{{\cal{N}}_2^{\mu \nu 
\alpha}}
{(k^2-m^2)^2[(k+p)^2-m^2]},
\ee
where
\be
{\cal{N}}_1^{\mu \nu
\alpha}=4i\left\{\left\{[(k+p)^2-m^2]k_{\beta}-2m^2p_{\beta}\right\}\epsilon^{\mu
\nu \alpha \beta}
-2p_{\sigma}k_{\beta}k^{\alpha}\epsilon^{\mu \nu 
\sigma
\beta}\right\} \,\, {\mbox{and}}
\ee
\be
{\cal{N}}_2^{\mu \nu
\alpha}=4i\left\{\left\{-[k^2-m^2](k+p)_{\beta}-2m^2p_{\beta}\right\}\epsilon^{\mu
\nu \alpha \beta}
-2p_{\sigma}k_{\beta}k^{\alpha}\epsilon^{\mu \nu 
\sigma
\beta}\right\}
\ee
Above we only considered the terms which do not vanish
after integration or because of symmetry properties in
the Lorentz indices. After 
some straightforward algebra, we can write
\be
I_1^{\mu\nu\alpha}=4i\left\{[J_{\beta}^{\Lambda}
-2m^2p_{\beta}J^1]
\epsilon^{\mu\nu\alpha\beta}-2p_{\sigma}g^{\alpha
\lambda}
\epsilon^{\mu\nu\sigma\beta}J_{\beta\lambda}^\Lambda
\right\}
\ee
and            
\be
I_2^{\mu\nu\alpha}=4i\left\{[-J_{\beta}^\Lambda
-2m^2p_{\beta}J^1-p_{\beta}J^\Lambda ]
\epsilon^{\mu\nu\alpha\beta}-2p_{\sigma}g^{\alpha
\lambda}
\epsilon^{\mu\nu\sigma\beta}J_{\beta\lambda}^{\Lambda}
\right\},
\ee
where we defined    
\bq 
J^1 = \int\frac{d^4k}{(2\pi)^4}
\frac{1}{(k^2-m^2)^2[(k+p)^2-m^2]}, 
\\
J^\Lambda = \int^\Lambda \frac{d^4k}{(2\pi)^4}
\frac{1}{(k^2-m^2)[(k+p)^2-m^2]}, \\
J_{\beta}^\Lambda = \int^\Lambda \frac{d^4k}{(2\pi)^4}
\frac{k_{\beta}}{(k^2-m^2)[(k+p)^2-m^2]}
\,\,\,\,\,\, \mbox{and} \\
J_{\beta\lambda}^\Lambda = \int^\Lambda
\frac{d^4k}{(2\pi)^4} \frac{k_{\beta}k_{\lambda}}
{(k^2-m^2)^2[(k+p)^2-m^2]}.
\eq
Among these integrals, the divergent are $J$,
$J_{\beta}$ and 
$J_{\beta\lambda}$. We
can manipulate them using (\ref{rr}) recursively to
obtain
\bq
J^\Lambda =I_{log}^\Lambda (m^2)-\tilde J, \\
J_{\beta}^\Lambda
=-2p^{\rho}\Theta_{\beta\rho}^\Lambda +\tilde
J_{\beta} \,\,\,\,\, 
\mbox{and} \\
J_{\beta\lambda}^\Lambda=\Theta_{\beta\lambda}^\Lambda-\tilde
J_{\beta\lambda},
\eq
where 
\bq
\Theta_{\alpha \beta}^\Lambda = \int^\Lambda
\frac{d^4k}{(2\pi)^4} \frac{k_\alpha
k_\beta}{(k^2-m^2)^3} \,\, , \\
\tilde J=\int\frac{d^4k}{(2\pi)^4} 
\frac{p^2+2p.k}{[k^2-m^2]^2[(k+p)^2-m^2]}\,\,
\mbox{and} \\
\tilde J_{\beta\lambda}=\int\frac{d^4k}{(2\pi)^4}
\frac{(p^2+2p.k)k_{\beta}k_{\lambda}}{[k^2-m^2]^3[(k+p)^2-m^2]}
\eq
Now that we removed the  external momentum dependence
from the divergent 
integrals, 
we note that they cancel out in $I_1^{\mu\nu\alpha}$
unambiguously. It remains
an undetermined finite term originated from a
difference between divergent integrals in
$I_2^{\mu\nu\alpha}$. Using that 
$p_{\beta}\tilde I=2\tilde I_{\beta}$ (which can be
shown by partial 
integration), we get
\be
I_1^{\mu\nu\alpha}=4i\left\{[\tilde 
J_{\beta}-2m^2p_{\beta}J^1]
\epsilon^{\mu\nu\alpha\beta}-2p_{\sigma}g^{\alpha
\lambda}
\epsilon^{\mu\nu\sigma\beta}\tilde
J_{\beta\lambda}\right\}
\ee
and
\be
I_2^{\mu\nu\alpha}=I_1^{\mu\nu\alpha}+\frac{\phi}{2
\pi^2}p_{\beta}
\epsilon^{\mu\nu\alpha\beta},
\ee  
where the divergent integrals were combined as in
(\ref{14}) to define
\be
\Upsilon^0_{\alpha \beta} \equiv g_{\alpha \beta}
I^\Lambda_{log}(m^2) - 4 \Theta_{\alpha \beta}^\Lambda
= \lambda g_{\alpha \beta}
\label{dec2}
\ee
in which $\lambda$ is a dimensionless, finite
parameter, and we have defined
\be
\phi \equiv \frac{8 \pi^2}{i} \lambda \, .
\ee
The finite integrals can be readily solved using
Feynman parameters after which we can write 
\be
\Pi^{\mu\nu\alpha}_{non-pert}= 
\epsilon^{\mu\nu\alpha\beta}\frac{p_{\beta}}{2 \pi^2}
\Bigg( \frac{\theta}{\sin \theta}- \phi \Bigg) \, ,
\label{jacre}
\ee  
where $\theta = 2 {\mbox{arcsin}}(\sqrt{p^2}/(2 m))$
and $p^2 < 4 m^2$. The equation above is similar to
the one encountered by Jackiw and Kosteleck\'y
\cite{JACKOS}, with
our $\phi$ playing the role of their surface term. Now
in order to arrive at their claimed unambiguous result
within the non-perturbative approach, another
information would have to be implemented. As it was
discussed in \cite{PEREZ} any regularization that had
broken the spherical symmetry in their explicit
integration would have altered their result. That is
the case of DR which breaks the tracelessness of the
combination $k_\mu k_\nu - 1/4 g_{\mu \nu}k^2$ in the
$4$-dimensional space-time. In calculating the surface
term which is originated from the shift in the
linearly divergent integral, one also makes  use
of such symmetry by performing a symmetric momentum
limit ${\mbox{lim}_{k \rightarrow \infty} \frac{k_\mu
k_\nu}{k^2} = \frac{g_{\mu \nu}}{4}}$. Therefore it is
an easy matter to check that if we use symmetric
integration in $\Theta_{\alpha \beta}$ in
(\ref{dec2}) then we will obtain that $\phi = 1/4$ in
(\ref{jacre}), to give in the limit of heavy fermion
mass the result found in
\cite{JACKOS},\cite{PEREZ},\cite{CHUNGJ}.

Now let us proceed to the perturbative in $b$
computation. The relevant diagrams are the
$b_\mu$-linear one loop correction to the photon
propagator in which a factor $i b_\lambda
\gamma^\lambda \gamma^5$ can be inserted in either of
the two internal fermionic lines to render equal
contributions. Thus the amplitude reads
\be
\Pi_{\mu \nu}^b = 2 \, (-i) b^\lambda \int \dq
{\mbox{tr}} \gamma_\mu S_F (k-p) \gamma_\nu S_F (k)
\gamma_\lambda \gamma^5 S_F (k) \equiv 2 b^\lambda
\Pi_{\lambda \mu \nu}\, ,
\ee
where $p$ is the external momentum. The integral above
is just our $I^{\mu \nu \alpha}_2$ in the
non-perturbative with $p \rightarrow -p$ and the
$\mu$, $\nu$ indices interchanged. Therefore we can
write, taking into account the change of signs,
\be
\Pi^{\mu\nu\alpha}_{pert}= 
\epsilon^{\mu\nu\alpha\beta}\frac{p_{\beta}}{2 \pi^2}
\left\{\frac{\theta}{\sin \theta}-\phi '\right\} \, ,
\ee 
The equation above is to be compared with equation
$(21)$ in the reference \cite{CHENDR}. Our
undetermined parameter $\phi '$ is just their ratio
$\ln(M1/M2)$. Thus we have achieved the same elegance
as it is expressed within differential regularization
with the advantage of working in the momentum space.
This result was expected since we have not made use of
an explicit regulator. We too have all the results
obtained in other regularization schemes embodied in
different choices for the parameter $\phi '$ which is
to be fixed on physical grounds either by symmetry
requirements or a renormalization condition; all the
possible ambiguities are expressed in terms of our so
called consistency relations which we left arbitrary
until the final stage in this case. It is therefore
not surprising
that our approach achieved the same merits as those
claimed within differential regularization.

It is interesting to observe that the indeterminacy
expressed by 
our parameters $\phi$ and $\phi '$ are ultimately
related to a non-vanishing
value for the CR. In other words the amplitudes
considered in this section 
are not independent of the momentum routing in the
loop. Would we have momentum routing independence then
the parameter $\lambda$ and consequently $\phi$ and
$\phi '$
would be zero. Generically we can state that the
presence of an axial vertex has broken such momentum
routing independence. In fact this is not a new
feature. It is
well known that only a special routing of the
integration momenta may result in a gauge invariant
answer in the presence of axial vertices
\cite{JACKOS}, \cite{JACTREI}. Notice also that
total vacuum polarization amplitude (\ref{tva})has a
contribution $\Pi_{\mu\nu}^0$, which corresponds to
pure $QED_4$. Would a particular routing choice 
violate gauge invariance in $\Pi_{\mu\nu}^0$? As we
have seen, if an arbitrary 
momentum routing is taken, then the $\Upsilon$'s must
be zero. On the other hand, it was shown in the last
section that if we choose a particular routing, there
is another possibility of mantaining gauge invariance
namely by fixing the relative coefficients of one
CR. Therefore, we can fix
the momentum routing in $\Pi_{\mu\nu}^b$ and then
adjust $c
= 2a$ and
$\Upsilon^2_{\mu\nu}=0$, so as to respect gauge
invariance. 

\section{Topological mass generation in
$3$-dimensional gauge theory}
\label{QED3}
As we have seen in section \ref{CPT}, CS terms can be
induced by radiative quantum effects even if they are
not present as bare terms in the original Lagrangian.
In $3+1$-dimensional space-time such terms could be
induced by extending the fermionic sector of $QED$
with an explicitly Lorentz and CPT violating
axial-vector term.  In $2+1$ dimensions, however, such
topological terms can naturally appear at quantum
level without any extension in the classical
Lagrangian \cite{RED}.  Consider the $QED_3$
Lagrangian with fermions of mass $m$, $L=-\frac{1}{4}
F_{\mu \nu}F^{\mu \nu} + \bar{\Psi}(iD\hspace{-2.3mm}/
+ m)\Psi$. Now let us study the role played by a
radiatively generated CS term in the sense of giving a
mass for the gauge field \footnote{Please see
\cite{DUNNE} for a complete account on this matter}.
In \cite{ROTHE} it was shown that despite being all
gauge invariant one could classify 
a set of regularizations in two groups: one in which
the originally massless boson remained massless (such
as Pauli-Villars regularization) and another in which
it turned out to be massive (such as DR among others).
Here we revisit this problem in the light of our
framework. Consider an expansion in powers of $A$ of
the one loop effective action, namely
\be
\Gamma_{QED_3}[A] = \mbox{tr} \ln \mbox{det} (i
\partial\hspace{-2.3mm}/ -m) + \mbox{tr} \Bigg(
\frac{1}{i \partial\hspace{-2.3mm}/-m}
A\hspace{-2.3mm}/ \Bigg) + \frac{1}{2}\mbox{tr} \Bigg(
\frac{1}{i \partial\hspace{-2.3mm}/-m}
A\hspace{-2.3mm}/ \frac{1}{i
\partial\hspace{-2.3mm}/-m} A\hspace{-2.3mm}/\Bigg) +
\ldots \, .
\ee
For a induced mass term, the relevant contribution is
the one that is quadratic in $A$ which we write
generically as
\be
\Gamma^2[A] = \frac{1}{2} \int \dtp A^\mu (p) \Pi_{\mu
\nu}(p) A^\nu (p)
\ee
where $\Pi_{\mu \nu}$ is the usual vacuum polarization
tensor,
\be
\Pi_{\mu \nu} (p) = \int \dt \mbox{tr} \Bigg(
\gamma_\mu 
\frac{p\hspace{-2.3mm}/+k\hspace{-2.3mm}/+m}{(p+k)^2 -
m^2} \gamma_\nu 
\frac{k\hspace{-2.3mm}/+m}{k^2 - m^2} \Bigg) = \int
\dt \frac{\eta_{\mu \nu}}{((p+k)^2-m^2)(k^2-m^2)}\, ,
\ee
with 
\be
\eta_{\mu \nu} = 2 \Big( 2 k_\mu k_\nu + 2 p_\mu k_\nu
- g_{\mu \nu} ( (k+p)\cdot k -m^2) + i m \epsilon_{\mu
\nu \alpha}p^\alpha \Big) \, ,
\ee
where we used that in $3$ dimensions $\mbox{tr}
\gamma^\mu \gamma^\nu \gamma^\alpha = - 2 i
\epsilon^{\mu \nu \alpha}$. Thus we can write
\be
\Pi_{\mu \nu}= 2 \Big( 2 I_{\mu \nu} + 2 p_\mu I_\nu +
i m \epsilon_{\mu \nu \alpha}p^\alpha I - g_{\mu
\nu}I^{(1)} -  g_{\mu \nu}p^\alpha I_\alpha \Big)\, ,
\ee 
where
\be
I,\, I_\mu \, , I_{\mu \nu} \equiv \int^\Lambda \dt
\frac{1, k_\mu , k_\mu k_\nu}{((p+k)^2-m^2)(k^2-m^2)}
\,\, \mbox{and}\,\, I^{(1)}\equiv \int^\Lambda \dt
\frac{1}{(p+k)^2-m^2} \, .
\ee
Among the integrals defined above only $I$ and $I_\mu$
are finite whereas the others can be rewritten with
(\ref{rr}) as
\be
I_{\mu \nu} = \int^\Lambda \dt \frac{k_\mu
k_\nu}{(k^2-m^2)^2} - \int^\Lambda \dt \frac{(p^2 + 2
p \cdot k) k_\mu k_\nu}{((p+k)^2-m^2)(k^2-m^2)}
\ee
in which the second integral on the RHS is finite and
\be
I^{(1)} = I_{lin}^\Lambda (m^2) - \int^\Lambda \dt
\frac{p^2+2p \cdot k}{((p+k)^2-m^2)(k^2-m^2)} ,
\label{i1}
\ee
with
\be
I_{lin}^\Lambda (m^2) \equiv \int^\Lambda \dt
\frac{1}{k^2-m^2} \, .
\ee
However the second integral on the RHS of (\ref{i1})
vanishes and therefore $I^{(1)} = I_{lin}^\Lambda
(m^2)$. With the results given above, $\Pi_{\mu
\nu}(p)$ reads:
\be
\Pi_{\mu \nu}(p) = 2 \Bigg( \Xi^1_{\mu \nu} + F_1
(p^2,m) \epsilon_{\mu \nu \alpha}p^\alpha + F_2(p^2,m)
\Big(\frac{p_\mu p_\nu}{p^2} - g_{\mu \nu} \Big)
\Bigg)\, ,
\ee
where $\Xi^1_{\mu \nu}$ is given as in (\ref{11}).
Generically we can write $\Xi^1_{\mu \nu} = \lambda 
g_{\mu \nu}$, on Lorentz invariance grounds where
$\lambda$ is a parameter (with dimension of mass) to
be determined. In order to assure gauge invariance, we
are led to set $\Xi^1_{\mu \nu}=0$ in this case
\footnote{A  linearly divergent term $\propto \Lambda
g_{\mu \nu}$ which would appear using a explicit
cut-off calculation \cite{JACKDES}does not appear in
our case as it would not appear in any gauge invariant
regularization such as DR.}. This  appears to be a
natural choice as no parity violating objects appear
in the vertex .  Moreover the finite coefficients
$F_1$ and $F_2$ evaluate to:
\be
F_1 (p^2,m) = \frac{i}{4 \pi} \Bigg[
\frac{m}{\sqrt{p^2}} \ln \Bigg(
\frac{1 + \sqrt{p^2/4m^2}}{1 - \sqrt{p^2/4m^2}} \Bigg)
\Bigg] \, ,
\ee
\be
F_2 (p^2,m) = \frac{1}{4 \pi} \Bigg[ m - \frac{1}{4
\sqrt{p^2}}(p^2 + 4m^2) \ln \Bigg( \frac{1 +
\sqrt{p^2/4m^2}}{1 - \sqrt{p^2/4m^2}} \Bigg)\Bigg] \,
.
\ee 
which is just the result that is obtained in DR
\cite{ROTHE},\cite{RED},\cite{JACKDES},\cite{GEGE} .
In the limit where $m \rightarrow \infty$ we obtain
\be
\Pi^{\mu \nu}_{m \rightarrow \infty}= \frac{i}{4
\pi}\frac{m}{|m|}\epsilon^{\mu \nu \alpha}p^\alpha \,
,
\ee
which contributes to the one loop effective action
with a term that in the coordinate space reads:
\be
\Gamma^2_{CS} = -\frac{i}{2}\frac{1}{4 \pi}\int d^3 x
\,\, \epsilon^{\mu \nu \alpha}A_\mu \partial_\nu
A_\alpha .
\ee
These results can be readily generalized to the
non-abelian case.

\section{Schwinger model and its chiral version}
\label{QED2}

As an example in $2$ dimensions, we study the
Schwinger model (ScM)($QED_2$ with massless
fermions)and its chiral version (CScM). The ScM is
exactly solvable \cite{MANTON} and has served as a
good laboratory for both testing theoretical
techniques and getting some insight in the vacuum
structure of $QCD_4$. Several non-trivial features of
the ScM and its massive and chiral version (such as
massive physical states formed via chiral anomaly,
instanton-like vacuum configurations labeled by a
$\theta$-angle, etc.) have counterparts in more
realistic theories \cite{SMAPPLIC}.
In the ScM, the massless photon of the tree
approximation acquires the mass $e^2/\pi$ ($e$ is the
coupling constant) at the one loop level (which is
exact in this case). Consider the effective action
radiatively induced by fermions:
\be
\Gamma_{S} = -i \ln \mbox{det}(i
\partial\hspace{-2mm}/ - e A\hspace{-2mm}/) \,.
\label{ea-scm}
\ee
The mass generation is seen at order $A^2$ which for
this model determines (\ref{ea-scm}) completely. Hence
all we need to do is to compute the vacuum
polarization tensor
\be
\Pi_{S}^{\mu \nu} (p) = i \mbox{tr} \int \dd \,
\gamma^\mu \frac{i}{k\hspace{-2mm}/}\gamma^\nu
\frac{i}{k\hspace{-2mm}/+p\hspace{-2mm}/}
\label{pmnir}
\ee
After taking the traces, using (\ref{rr}) to write the
divergence as a function 
of the loop momentum only and evaluating the finite
integrals, we obtain (see also \cite{JACKIWFU}):
\be
\Pi_{S}^{\mu \nu} (p) = \Pi^{\mu \nu}_{\infty} +
\frac{1}{\pi}\Big(\frac{g^{\mu \nu}}{2} - \frac{p^\mu
p^\nu}{p^2}\Big)
\label{pmnsm}
\ee
where
\be
\Pi^{\mu \nu}_{\infty} \equiv 2 i \int^\Lambda \dd
\frac{(-k^2 g^{\mu \nu} + 2 k^\mu k^\nu)}{(k^2 -
\mu^2)^2}
\label{pmninf}
\ee
and $\mu^2$ is an infrared cut-off which is immaterial
for the value of $\Pi^{\mu \nu}_{\infty}$. Some
features are noteworthy. Notice that in general
$\Pi_{S}^{\mu \nu}$ is not gauge invariant. Lorentz
invariance tell us that $\Pi^{\mu \nu}_{\infty}$
should be proportional to
to $g_{\mu \nu}$ but the coefficient is in principle
undetermined since the integral is divergent.  
Moreover if $\Pi^{\mu \nu}_{\infty}$ assumes any value
different from zero it would break the traceless of
$\Pi_{S}^{\mu \nu}$ already manifest in its integral
representation (\ref{pmnir}). However Pauli-Villars or
DR can be employed and gauge invariance  restored
within this schemes. DR, for instance, evaluates
(\ref{pmninf})
to $\frac{1}{2 \pi} g^{\mu \nu}$ which gives for
(\ref{pmnsm}):
\be
{\overline{\Pi}}_{S}^{\mu \nu} (p) =
\frac{1}{\pi}\Big( g^{\mu \nu} - \frac{p^\mu
p^\nu}{p^2}\Big)
\label{pmnsmgi}
\ee
Now recall the CR (\ref{11A}), namely 
\be
\Delta_{\mu \nu}^0 = \int^{\Lambda}
\frac{d^2k}{(2\pi)^2}\frac{g_{\mu\nu}}{k^2-m^2}-2\int^{\Lambda}
\frac{d^2k}{(2\pi)^2}\frac{k_{\mu}k_{\nu}}{(k^2-m^2)^2},
\ee
The  choice $\Delta_{\mu \nu}^0 = 0$ can be used in
(\ref{pmninf}) to obtain a  result which is just 
$\frac{1}{2 \pi} g^{\mu \nu}$ as it can be easily
demonstrated. Hence gauge invariance is restored
within our framework. This is close in spirit to
Jackiw's
approach in \cite{JACKIWFU}. In other words, we can
state that this particular value for  $\Delta_{\mu
\nu}^0$ is the one which restores gauge invariance, if
we wish so. It plays the role of an undetermined local
part in the quadratic term of the effective action. If
the ScM really described a physical particle, we could
say that we had to choose $\Delta_{\mu \nu}^0$ to
vanish so to explain the photon mass $m^2 = e^2/\pi$.

In order to gain some more intuition, let us make a
similar analysis with the CScM.
We simply substitute the vector interaction with a
chiral interaction in (\ref{ea-scm}):
\be
\Gamma_{\chi } = -i \ln \mbox{det}(i
\partial\hspace{-2mm}/ - e
(1+\gamma_5)A\hspace{-2mm}/) \, ,
\label{ea-cscm}
\ee
$\gamma_5 = \gamma_0 \gamma_1$. An analogous
calculation leads us to the result
\be
\Pi_{\chi }^{\mu \nu} (p) = \Pi_{S}^{\mu \nu} (p) +
g_{\alpha \beta} \Big( \epsilon^{\nu
\alpha}\Pi_{S}^{\mu \beta} (p) + \epsilon^{\mu
\alpha}\Pi_{S}^{\beta \nu} (p) \Big) + \epsilon^{\mu
\alpha}\epsilon^{\nu \beta}\Pi_{S \,\, \alpha \beta}
(p) \, ,
\label{pmn-cscm}
\ee
where we used that
\be
\gamma_5 \gamma_\mu = \epsilon^{\mu \nu} \gamma_\nu \,
,
\label{gameps}
\ee
and $\Pi_{S}^{\mu \nu} (p)$ is given as in
(\ref{pmnsm}). As it is well known \cite{JACKIWRAJA}
there occurs a chiral anomaly in this model: it cannot
be made gauge invariant. This is a manifestation of
the anomalous non-conservation of the chiral current
in the ScM for 
\be
p_\nu {\Pi}^{\mu \nu}_{5} = -
\frac{1}{\pi}\tilde{p}^\mu  \rightarrow \partial^\nu
j_\nu^5 = \frac{e}{\pi} \epsilon_{\nu \mu}
\partial^\nu A^\mu \, ,
\label{awi}
\ee
where $ {\Pi}^{\mu \nu}_{5} = \epsilon^{\nu
\kappa}({\overline \Pi}^\mu_\kappa)_S$ because of
(\ref{gameps})and $\tilde{p}^\nu = \epsilon^{\nu
\alpha}p_\alpha$ \, \footnote{Notice that
$(\overline{\Pi}_\mu^\mu)_{S}= 1/\pi$ which provides
precisely the value of the anomaly.}. 

Now let us write generically for the CR 
$$
\Delta_{\mu \nu}^0 = \frac{\lambda }{2 \pi} g_{\mu
\nu}
$$ 
based on Lorentz invariance ($\lambda$ is a
dimensionless parameter). Thus 
\be
\Pi^{\mu \nu}_{\infty}  = \Bigg( \frac{\lambda + 1}{2
\pi}  \Bigg) g^{\mu \nu}
\ee
from which we see that the choice $\lambda=0$ enforces
gauge invariance on the ScM.
We can rewrite the axial Ward identity (\ref{awi}) as
a function of $\lambda$, namely
\be
p_\nu {\Pi}^{\mu \nu}_{5} = - \frac{\lambda + 2}{2
\pi} \,\, \tilde{p}^\mu \, .
\label{awil}
\ee
Had we opted for preserving the AWI, we would have to
set $\lambda = -2$. This in turn would transfer the
anomaly to the VWI since
\be
p_\mu \Pi^{\mu \nu}_S \Big|_{\lambda = -2} = -
\frac{1}{\pi}p^\nu \, ,
\ee
as expected.

On the other hand, for the CScM, (\ref{pmn-cscm})
yields 
\be
\Pi_{\chi }^{\mu \nu} (p) = \frac{1}{\pi} \Bigg(
(\lambda + 2)g^{\mu \nu}
-  (g^{\mu \alpha}+\epsilon^{\mu \alpha})
\frac{p_\alpha p_\beta}{p^2} (g^{\beta \nu}-
\epsilon^{\beta \nu}) \Bigg) \, .
\ee
Unlike the ScM, imposing gauge invariance does not fix
the value of $\lambda$ since
\be
p_\mu \Pi_{\chi }^{\mu \nu} (p) = \frac{1}{\pi}\Big(
(\lambda +1) p^\nu - \tilde{p}^\nu \Big) \, ,
\ee
which shows that the longitudinal part does not vanish
for any value of $\lambda$.
Despite the lack of gauge invariance and the arbitrary
parameter $\lambda$, it constitutes a perfect sound
theory \cite{JACKIWRAJA}. It can be exactly solved to
find that for $\lambda > - 1$ it is a unitary and
positive definite  model, in which the photon acquires
a mass
\be
m^2 = \frac{e^2}{\pi} \frac{(\lambda +2)^2}{\lambda +
1}
\ee
An equivalent formulation in a bosonized version of
the CScM places  $\lambda$ as arising from ambiguities
in the bosonization procedure. In fact the CScM can be
formulated in a gauge invariant way in which a
Wess-Zumino term \footnote{Such term arise naturally
by adopting the Faddeev-Popov trick for quantising a
theory with an anomaly.} exactly cancels the variation
of the original Lagrangian under a gauge
transformation \cite{HARADA}. In addition, it was
shown \cite{KRAMER} that the anomalous formulation is
nothing but a special gauge (unitary gauge $g=1$) of
the gauge invariant formulation. Had we chosen the
value $\lambda =0$ as we did for the ScM we would
obtain $m^2 = \frac{4 e^2}{\pi}$. Curiously this value
has already been conjectured within another
regularization (Faddeevian regularization)
\cite{MITRA}; however it turned out to be a special
case  of the CScM with a minimal Wess-Zumino term with
a  restriction on an undetermined parameter
correspondent to $\lambda$ \cite{FR=SMWZ}.

It is important to remark that there is no reason to
impose $\lambda =0$ for
the CScM as we did for the ScM. The best we could do
based on unitarity and positivity of the theory was to
establish a range of values for $\lambda$. This
remains true in its gauge invariant formulation since
it obviously yields the same induced mass for the
photon.
Within our framework, we can somewhat generalise the
ideas proposed by 
Jackiw \cite{JACKIWFU} in the treatment of the ScM and
the CScM to perturbative calculations in any quantum
field theory where ultraviolet divergences appear. The
latter can always be displayed either by basic
divergent integrals or by differences between
integrals of the same degree of divergence
whose value is finally fixed by imposing vital
symmetries from the theory and/or by fitting with
experimental data.

\section{Concluding remarks and outlook}

In this paper we extended an implicit regularization
scheme to be applicable in quantum field theories
defined in $n$ space-time dimensions. As we do not
leave the integer dimension in which the theory is
defined, parity violating objects present in chiral or
topological field theories need not to be
dimensionally continued and therefore we avoid  well
known ambiguities involved in this procedure. Moreover
all the undeterminacies will be cast into a set of CR
to be fixed on  physical grounds either by imposing
that the vital symmetries must not be violated or by
experiment.

In this sense our framework is useful to simplify loop
calculations in, for instance, Chern-Simons-Matter
theories \cite{KUNST}. That is because
High-Covariant-Derivative regularizations turn the
calculations extremely lengthy (especially beyond one
loop order) due to the complicated form that the gauge
field propagator assumes. 
Moreover there are cases where it seems to be possible
to opt for a HCD or an extended DR; in  \cite{WUCHEN2}
the one loop shift in noncommutative CS coupling 
depends on this choice. Therefore even if one uses
different regularizations which respect fundamental
symmetries of a theory (such as gauge invariance) one
may not get the same radiative correction. This is
different from the situation when a theory possesses
an intrinsic ambiguity whose value may have to be
fixed only by experiment,
even if the renormalization is finite \cite{JACKIWFU}.
As our framework does not modify or corrupt the
underlying theory in consideration, it constitutes an
ideal tool to study these problems.

Should any further constraint be imposed, such as
(re)normalization conditions or some other physical
requirement, they can be readily implemented within
our framework \cite{BON}. Our main concern within this
formulation was to keep the ambiguities to be fixed in
the very final stage of the calculation. 

When overlapping divergences occur, they are treated
in a similar fashion
\cite{AOMC}. Finally our approach may be generalized
to multi-loop calculations. The proof follow the same
lines as the forest and skeleton construction in the
BPHZ formulation \cite{WIP}. 

\section*{Acknowledgments}
The authors wish to thank Dr. J.L. Acebal, Dr.
Brigitte Hiller, Dr.
Olivier Piguet, Dr. M.O.C.Gomes
and Dr. A.L.Mota for fruitful discussions throughout
the development of this work. This work is supported
by PRAXIS XXI-BPD/22016/99-FCT-MCT (Portugal), FAPEMIG
(Brazil) and CNPq (Brazil).

\end{document}